# Characterizing the Reliability of a Novel Upright CT for Proton Therapy

Yuhao Yan, M.Sc.,[*, †] Jordan M. Slagowski, Ph.D.,[*] Jessica R. Miller, Ph.D.,[*] John W. Hayes, Ph.D.,[‡] Carson Hoffman, Ph.D.,[‡] Minglei Kang, Ph.D.,[*] Carri K. Glide-Hurst, Ph.D.[*, †]

[*]*Department of Human Oncology, University of Wisconsin-Madison, Madison, WI, USA*
[†]*Department of Medical Physics, University of Wisconsin-Madison, Madison, WI, USA*
[‡]*Leo Cancer Care, Middleton, WI, USA*

**CORRESPONDENCE**
Carri K. Glide-Hurst, Ph.D., Department of Human Oncology, University of Wisconsin-Madison, 600 Highland Avenue, Madison, WI 53792.
Email: glidehurst@humonc.wisc.edu

**ACKNOWLEDGMENTS**
Work reported in this publication was supported in part by the National Cancer Institute of the National Institutes of Health under award numbers: R01CA204189 and R01HL153720 (PI: Carri Glide-Hurst). The content is solely the responsibility of the authors and does not necessarily represent the official views of the National Institutes of Health.

**CONFLICT OF INTEREST STATEMENT**
Carri Glide-Hurst reports research collaborations with RaySearch Laboratories and Leo Cancer Care outside of the submitted work (PI: Carri Glide-Hurst). John Hayes and Carson Hoffman are employees of Leo Cancer Care. Yuhao Yan, Jordan Slagowski, Jessica Miller and Minglei Kang report no conflict of interests.

**DATA AVAILABILITY STATEMENT**
Research data are not available at this time.

## Abstract

**Background:** The novel upright patient positioning systems enable upright proton therapy yet the reliability of the coupled upright CT to support adaptive proton therapy needs assessment.

**Purpose:** To evaluate longitudinal stability and reliability of upright CT for proton dose calculation and feasibility of a simplified phantom configuration for accelerated routine QA.

**Methods:** A calibration phantom with tissue mimicking inserts was scanned for 14 imaging sessions on an upright CT following consensus guidelines over 7 months. Intersession repeatability of CT number was assessed by standard deviation (SD). Hounsfield look-up table (HLUT) of stopping power ratio (SPR) was derived for proton dose calculation. Body- and head-phantom derivations were compared to assess size dependency. The proposed simplified phantom configuration using a different insert arrangement was scanned for 15 imaging sessions over 8 months. Inter- and intra-session repeatability were assessed. CT numbers and SPR derivations were compared with consensus configuration results. Both configurations were scanned on a conventional recumbent CT for a single session to validate the findings. An anthropomorphic phantom was scanned on upright and recumbent CT. Targets were drawn mimicking spine and prostate disease. Proton plans were developed on upright CT for each site using pencil beam scanning techniques, robust optimization (3-5 mm setup uncertainty, 3.5% range uncertainty) and Monte Carlo algorithm, and applied to co-registered recumbent CT. Equivalence of dose calculation using different calibrations (consensus vs simplified) on different CT machines were assessed via controlled comparisons. Imaging protocol was fixed for different phantoms and CT machines throughout the experiment.

**Results:** Simplified configuration measured CT numbers of all inserts in one scan while five were needed following consensus guidelines. For both phantom configurations, upright CT demonstrated excellent longitudinal CT number stability with minimal inter- and intra-session SD (<4.9 HU and <1.6 HU, respectively). System upgrades and recalibration introduced marginal offset (<9 HU). Size dependency was identified with statistically significant ($p<0.05$) differences in upright CT numbers, propagated to ΔSPR up to 5.3% comparing body and head phantom results. Statistically significant ($p<0.05$) differences were also found comparing upright CT numbers measured by two configurations with ΔSPR up to 2.6%. Recumbent CT assessment suggested smaller ΔSPR (maximum 0.7%). For both spine and prostate proton plans, comparison suggested local dose differences up to 8% of prescription dose (greatest when comparing upright CT dose calculated using different phantom configuration), while clinical equivalence was found with target coverage differences <0.2% and gamma pass rates=100% at 3 mm/3% for all controlled comparison of different CT machines and phantom configurations.

**Conclusions:** The upright CT demonstrated excellent longitudinal stability and reliability to support proton ART. The simplified configuration shows feasibility for rapid QA to monitor machine stability. Multi-institution validation is warranted.

## Introduction

With the introduction of novel upright patient positioning systems, there has been an increasing interest in delivering radiation therapy (RT)—particularly particle therapy—with patients in the upright orientation (e.g., seated, perched) by rotating patients relative to a fixed beamline. By eliminating the need of a large rotating gantry, upright particle therapy allows small footprint of the treatment room and easier construction and maintenance of the delivery system, reducing the cost and offering potential improvements to the accessibility of particle therapy.[1,2] In addition, when patients are positioned upright, potential anatomical benefits may be introduced such as increasing lung volume, reducing tumor motion in the lung[3,4], and stabilizing anatomy in the pelvis,[5] potentially leading to better normal tissue sparing.

To support upright RT, imaging of patients in the upright orientation is necessary for simulation, treatment planning, and daily image guidance. Novel imaging systems have been introduced where the CT gantry/ring translates vertically relative to the patient,[6–8] contrary to conventional recumbent CT scanners where the gantry is fixed while the patient is translated by the couch. One distinct benefit of integrated vertical CTs at the treatment isocenter is that uncertainty is reduced as the patient does not move between imaging and treatment isocenters, thereby minimizing anatomical variations and uncertainties.[1,9] Isocentric upright CT enables rapid acquisition of high quality CT images which is an essential step to support online ART. However, the longitudinal stability of CT numbers is critical to enable accurate and consistent dose calculations on planning and daily CT images while utilizing the same calibration curve,[10,11] as it has been reported that longitudinal variation in CT number translates to variations in stopping power ratios (SPRs) of similar scale.[12]

Recently, one of the first clinical installations of a novel isocentric upright CT scanner coupled with a 6 degree-of-freedom upright patient positioner was performed for future integration with a fixed proton therapy beam line. This work sought to quantify the longitudinal repeatability of CT number over 8 months of operation. Considerations were given to derive the Hounsfield look-up table (HLUT) between CT number and SPR following consensus guidelines[13], and a simplified phantom configuration was assessed to investigate quantitative and dosimetric differences to the consensus configuration to investigate its feasibility for performing routine QA on both the upright and recumbent CT.

## Methods

The Marie upright CT (Serial Number 6, Leo Cancer Care, Middleton, WI) is demonstrated in Figure 1A. The CT system is comprised of three major components: the pillars that support the gantry and execute the tilt of ±15º, the gantry that supports the CT ring with associated translational hardware, and the 85 cm diameter CT ring itself that translates vertically within a range of 180 cm. The CT is coupled with a 6 degree-of-freedom patient positioning system (chair). The reconstruction field-of-view (FOV) is up to 62.4 cm. The tube is designed to operate within 80-140 kVp and up to 250 mAs, while only 120 kVp was currently calibrated for our machine.

To characterize the upright CT for proton dose calculations, an Advanced Electron Density Phantom (Sun Nuclear, Melbourne, FL) was used which included tissue-mimicking inserts (lung, adipose, soft tissue and bone) with known density and chemical compositions. The CT calibration followed established consensus guidelines as described by Peters *et al.*[13] For each calibration session, 14 scans were consecutively acquired including: (1) five scans to measure insert CT numbers with the body phantom, where one initial scan was acquired with all non-dense inserts placed in the phantom, and four additional scans were acquired, each with the central insert replaced with one of the four bone inserts (nominal mass density 1.21-1.93 g/cm$^3$) to reduce uncertainty from beam hardening, (2) four scans to evaluate the potential impact of beam hardening due to insert location for dense inserts, each with one fixed periphery insert replaced with one of the four bone inserts, and (3) five scans to measure insert CT numbers with the head phantom following the same procedure as (1). An example of phantom configuration is shown in Figure 1B-C. All scans were performed using an identical protocol of 120 kVp, 250 mAs, 1 pitch, 1 s per rotation, ~1 mm$^2$ in-plane resolution, 500 mm$^2$ FOV, 2 mm slice thickness, and a vendor-supplied RamLak reconstruction kernel using filtered backprojection. CT numbers were automatically extracted using a Python algorithm developed based on Pylinac,[14] where a cylindrical region of interest (ROI, 8 mm radius and 10 slices/20 mm height) was automatically identified in the center of each insert, and mean HU of each ROI was calculated.

To assess the longitudinal stability of the CT numbers, 14 imaging sessions were performed over a span of 7 months (Jan-July 2025). Intersession repeatability was assessed by calculating the standard deviation (SD) and coefficient of variation ($CV = \frac{SD}{mean}$) of the 14 sessions using the body phantom data. During the course of the study, the upright CT underwent software upgrades twice, with a recalibration of the water CT number using an ACR 464 phantom (Sun Nuclear) during one software upgrade. Absolute and percentage differences for each timepoint were calculated referring to the initial timepoint as the baseline.

The Hounsfield look-up table (HLUT) of stopping power ratio (SPR) was further derived for proton dose calculation using an algorithm provided by the consensus guidelines[13]. SPR of each insert was computed following the Bethe equation in combination with mean excitation energies from ICRU-49.[15] Tabulated human tissues[16] were also included to improve the calibration stability where CT number estimation[17] and SPR calculation was performed. After grouping by four tissue types (lung, adipose, soft tissue and bone), piecewise linear regression was performed between CT numbers and SPRs to derive the HLUT.

Per consensus guidelines, body and head-specific HLUTs were derived individually following above methodology. CT numbers and HLUTs acquired with the body and head phantoms were compared to assess the potential size-dependency of beam hardening and the need of site-specific HLUT. To test for statistically significant differences between CT number quantification using the body and head phantom, Wilcoxon signed-rank test was performed using the pair data from the 14 sessions, with p-value <0.05 considered statistically significant difference.

In addition to the consensus measurements, an alternate phantom configuration was assessed using a single insert arrangement to simplify the acquisition process[18]. As the upright scanner is expected to support online adaptive radiation therapy procedures, establishing efficient means of performing routine CT QA is essential. Thus, a simplified phantom configuration (termed "simplified" in this work) was scanned with all four bone inserts simultaneously placed in the body phantom (Figure 1D-E) such that CT numbers of all inserts could be extracted using a single scan, contrary to the consensus configuration where five scans were needed to measure CT numbers of all inserts. The simplified configuration was scanned across 15 sessions over a span of 8 months (Dec 2024-July 2025) with five repeat acquisitions performed at each session. During 14 of the sessions, datasets were also acquired using the consensus configuration. For the simplified configuration, the mean and SD across the five repeat acquisitions were calculated for each session. Intrasession and intersession repeatability were calculated and HLUT was derived following the methods for the consensus configuration. CT numbers and HLUT were compared with the consensus results by calculating absolute and percentage differences. To test for statistically significant differences between CT number quantification following the consensus and simplified configurations, Wilcoxon signed-rank test was performed using the data from the 14 matched sessions, with p-value <0.05 considered statistically significant difference.

As a comparator to the standard of care (e.g., recumbent CT), a single session with both the consensus and simplified phantom configurations were performed on a Siemens SOMATOM Definition Edge (Siemens Healthineers, Erlangen, Germany) using a matched imaging protocol with the exception of the reconstruction kernel (Br38s, a vendor-specific standard kernel for body imaging). CT numbers were extracted and HLUT was derived following the same methodology described above. Results from different configurations were compared by calculating absolute and percentage differences.

An anthropomorphic phantom (ATOM, Sun Nuclear) was scanned on the upright CT (120 kVp, 570 mm$^2$ FOV, ~ 1.1 mm$^2$ in-plane resolution). Tumor-mimicking targets were delineated to represent the spine and prostate/seminal vesicles. Proton treatment plans were developed in the RayStation Treatment Planning System (v2024A SP3, Stockholm, Sweden) for each target using pencil beam scanning techniques and a Hitachi ProBeat beam model. The spine plan included three beams (posterior-anterior, left posterior-oblique (50º) and right posterior-oblique (50º)) prescribed to 9 Gy (RBE)/fx for 3 fractions to target $D_{95}$, which was robustly optimized using 3 mm setup uncertainty and 3.5% range uncertainty. The prostate plan included two opposing beams (right lateral and left lateral) prescribed to 8 Gy (RBE)/fx for 5 fractions to target $D_{98}$, which was robustly optimized using 5 mm setup uncertainty and 3.5% range uncertainty following standard clinical planning guidelines. Dose was calculated using the body HLUT derived from the consensus guidelines and evaluated on the same dataset but using the HLUT from the simplified phantom configuration for comparison to be implemented for routine QA. Monte Carlo v5.6 dose engine was used with 0.5% statistical uncertainty.

To verify the feasibility of the simplified configuration and determine agreement with standard of care recumbent CT, the phantom was scanned on the recumbent CT using a matched protocol with the exception of the vendor-specific reconstruction kernel. The recumbent CT dataset was rigidly registered to the upright CT dataset, and the plan was copied and recalculated on the recumbent CT dataset using the body HLUTs derived from both configurations.

To evaluate dosimetric differences, target coverage was compared, and global 3D Gamma analysis was performed at 3 mm/3% criterion with 10% dose threshold for three dose evaluation comparisons:
   (1) equivalence between HLUT calibration using the consensus and simplified phantom configuration on upright CT,
   (2) equivalence between HLUT calibration using the consensus and simplified phantom configuration on recumbent CT,
   (3) equivalence of upright and recumbent CT dose calculation using the HLUT calibrated following the consensus configuration.

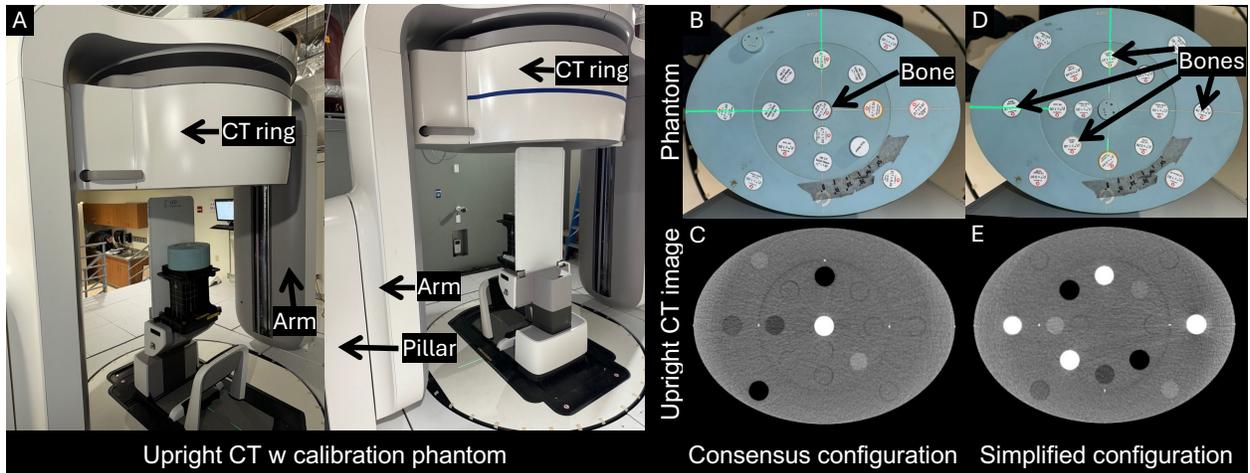

**Figure 1.** Demonstration of (A) the upright CT and the calibration phantom setup, along with different insert configurations and corresponding upright CT images including (B-C) consensus (Peters *et al.*, 2023) and (D-E) simplified configuration.

# Results

Figure 2 summarizes the longitudinal stability of upright CT numbers extracted from the scans acquired with the consensus or simplified configurations using the body phantom with software upgrades and recalibration timepoints noted. Supplement Table S1 summarizes longitudinal statistics for each insert scanned at upright CT with different phantom configurations. Overall, excellent repeatability of upright CT numbers was observed for all inserts with intersession SD<4.9 HU when scanned following the consensus configuration. Similarly, for the simplified configuration, excellent repeatability was observed with intersession SD <3.2 HU and intrasession SD <1.6 HU. For both consensus and simplified configurations, inter- and intra-session CVs were within 8% for all cases except for few outliers that were inflated by the small mean values.

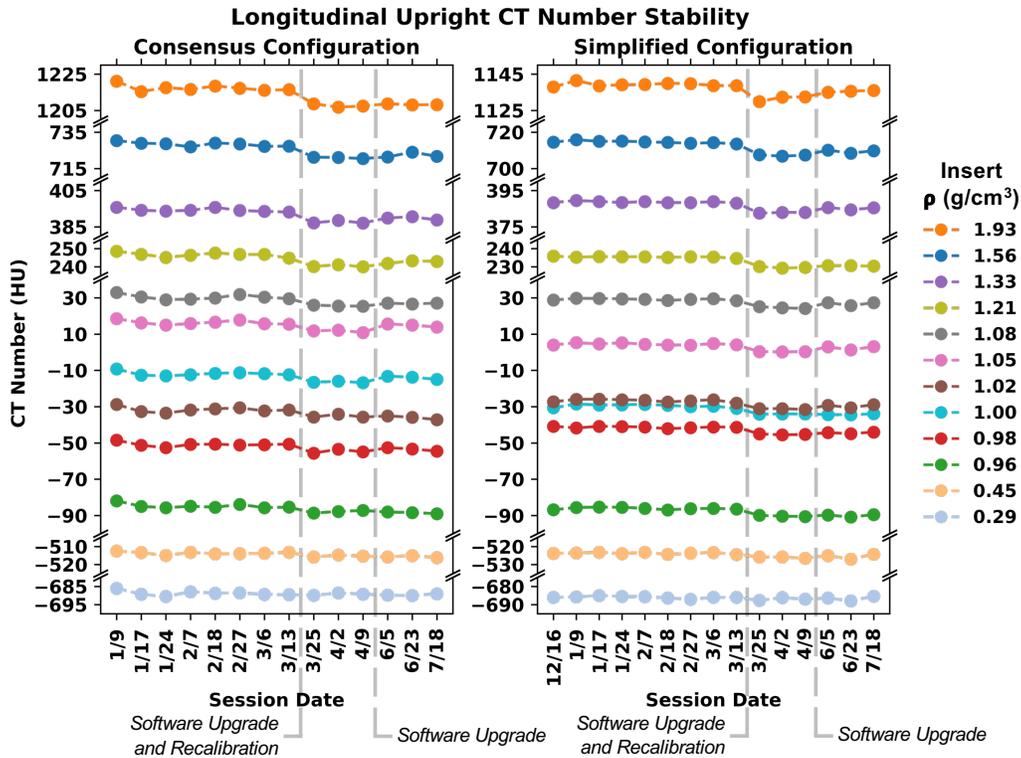

**Figure 2.** Longitudinal stability of upright CT number acquired following different phantom configurations using the body phantom

**Table S1.** Summarized longitudinal statistics for upright CT numbers of all inserts acquired using different phantom configurations with the body phantom. For intra-session metrics, range was reported (in absolute value in case the mean is negative). SD: standard deviation. CV: correlation of variation.

| Machine | Upright CT | | | | | | | |
|---|---|---|---|---|---|---|---|---|
| Configuration | Consensus | | | Simplified | | | | |
| Metric | Mean (HU) | Inter-session SD (HU) | Inter-session CV (%) | Mean (HU) | Intra-session SD (HU) | Intra-session CV (%) | Inter-session SD (HU) | Inter-session CV (%) |
| Lung LN300 | -689 | 1.1 | 0.2 | -686 | [0.4, 1.3] | [0.1, 0.2] | 0.8 | 0.1 |
| Lung LN450 | -514 | 1.1 | 0.2 | -525 | [0.5, 1.5] | [0.1, 0.3] | 1.2 | 0.2 |
| HE General Adipose | -86 | 2.0 | 2.3 | -88 | [0.5, 1.4] | [0.6, 1.5] | 2.1 | 2.3 |
| HE Breast 50:50 | -52 | 1.9 | 3.7 | -43 | [0.3, 1.2] | [0.7, 2.6] | 1.8 | 4.2 |
| Liquid Water | -13 | 2.1 | 15.6 | -31 | [0.4, 1.3] | [1.3, 4.3] | 2.4 | 7.5 |
| HE CT Solid Water | -33 | 2.3 | 6.9 | -28 | [0.6, 1.3] | [2.3, 4.7] | 2.0 | 7.1 |
| HE Brain | 15 | 2.1 | 14.1 | 3 | [0.5, 1.3] | [11.8, 1033.3] | 1.8 | 55.0 |
| HE Liver | 29 | 2.3 | 8.1 | 28 | [0.3, 1.4] | [1.1, 5.5] | 1.9 | 6.9 |
| HE Inner Bone | 244 | 2.8 | 1.1 | 233 | [0.3, 1.0] | [0.1, 0.4] | 2.6 | 1.1 |
| CaCO3 30% | 392 | 3.0 | 0.8 | 387 | [0.2, 1.6] | [0.1, 0.4] | 2.4 | 0.6 |
| CaCO3 50% | 726 | 3.5 | 0.5 | 712 | [0.4, 1.3] | [0.1, 0.2] | 3.2 | 0.5 |
| HE Cortical Bone | 1213 | 4.9 | 0.4 | 1137 | [0.2, 1.0] | [0.0, 0.1] | 3.2 | 0.3 |

Supplement Table S2 summarizes upright CT number deviations of each timepoint from the first data acquisition baseline following the consensus configuration with the body phantom, grouped by system upgrades. The selected baseline was systematically higher than all of the following timepoints. After the 1st system upgrade, the magnitude of mean ΔHU increased by up to 9 HU, especially for bone inserts. These offsets were persistent after the 2nd system upgrade in accordance with Figure 2.

**Table S2.** Longitudinal upright CT number changes with regard to the baseline (the first available timepoint) grouped by system upgrades, acquired following the consensus configuration using the body phantom.

| Timepoint | Baseline | Before 1st upgrade + recalibration (N=7) | | After 1st upgrade + recalibration (N=3) | | After 2nd upgrade (N=3) | |
|---|---|---|---|---|---|---|---|
| Metric | CT number (HU) | Mean Δ (HU (%)) | Δ range (HU (%)) | Mean Δ (HU (%)) | Δ range (HU (%)) | Mean Δ (HU (%)) | Δ range (HU (%)) |
| Lung LN300 | -686 | -3.1 HU (-0.4 %) | [-4.5, -1.8] HU ([-0.7, -0.3] %) | -3.2 HU (-0.5 %) | [-3.9, -2.5] HU ([-0.6, -0.4] %) | -3.5 HU (-0.5 %) | [-3.9, -3] HU ([-0.6, -0.4] %) |
| Lung LN450 | -513 | -1.3 HU (-0.2 %) | [-2.4, -0.7] HU ([-0.5, -0.1] %) | -2.8 HU (-0.5 %) | [-3.3, -2.2] HU ([-0.6, -0.4] %) | -3.1 HU (-0.6 %) | [-3.6, -2.5] HU ([-0.7, -0.5] %) |
| HE General Adipose | -82 | -3.3 HU (-4 %) | [-3.9, -2] HU ([-4.7, -2.5] %) | -6 HU (-7.3 %) | [-6.8, -5.4] HU ([-8.3, -6.5] %) | -6.6 HU (-8.1 %) | [-7.1, -6.2] HU ([-8.7, -7.6] %) |
| HE Breast 50:50 | -48 | -2.7 HU (-5.5 %) | [-4.1, -2.1] HU ([-8.5, -4.4] %) | -6.2 HU (-12.8 %) | [-7.1, -5] HU ([-14.7, -10.3] %) | -5 HU (-10.3 %) | [-6.1, -4.1] HU ([-12.5, -8.5] %) |
| Liquid Water | -9 | -2.9 HU (-31.3 %) | [-3.8, -2.1] HU ([-40.6, -22.1] %) | -7.2 HU (-77.4 %) | [-7.5, -6.7] HU ([-80.9, -72.3] %) | -4.7 HU (-50.7 %) | [-5.7, -4] HU ([-61.4, -42.7] %) |
| HE CT Solid Water | -29 | -3.2 HU (-11.2 %) | [-4.8, -1.9] HU ([-16.6, -6.5] %) | -6.4 HU (-22.3 %) | [-7, -5.4] HU ([-24.2, -18.6] %) | -7.2 HU (-25.2 %) | [-8.4, -6.3] HU ([-29.1, -21.9] %) |
| HE Brain | 18 | -2.4 HU (-13.1 %) | [-3.6, -0.8] HU ([-19.3, -4.1] %) | -6.9 HU (-37.4 %) | [-7.6, -6.4] HU ([-41.3, -34.6] %) | -3.7 HU (-20 %) | [-4.6, -2.9] HU ([-24.9, -15.9] %) |
| HE Liver | 33 | -2.9 HU (-8.7 %) | [-4, -1] HU ([-12, -3.1] %) | -7.3 HU (-22.2 %) | [-7.6, -6.9] HU ([-23, -21] %) | -6 HU (-18.3 %) | [-6.3, -5.9] HU ([-19.1, -17.9] %) |
| HE Inner Bone | 249 | -2.2 HU (-0.9 %) | [-3.8, -0.9] HU ([-1.5, -0.4] %) | -8.1 HU (-3.3 %) | [-8.5, -7.4] HU ([-3.4, -3] %) | -5.8 HU (-2.3 %) | [-6.7, -5.2] HU ([-2.7, -2.1] %) |
| CaCO3 30% | 396 | -1.7 HU (-0.4 %) | [-2.6, 0] HU ([-0.7, 0] %) | -8.2 HU (-2.1 %) | [-8.7, -7.3] HU ([-2.2, -1.8] %) | -6.1 HU (-1.5 %) | [-7, -5.2] HU ([-1.8, -1.3] %) |
| CaCO3 50% | 730 | -2.3 HU (-0.3 %) | [-3.5, -1.3] HU ([-0.5, -0.2] %) | -9.4 HU (-1.3 %) | [-9.9, -9.1] HU ([-1.4, -1.2] %) | -8 HU (-1.1 %) | [-9, -6.3] HU ([-1.2, -0.9] %) |
| HE Cortical Bone | 1221 | -4.3 HU (-0.4 %) | [-5.8, -2.7] HU ([-0.5, -0.2] %) | -13.6 HU (-1.1 %) | [-14.4, -12.5] HU ([-1.2, -1] %) | -12.9 HU (-1.1 %) | [-13.1, -12.5] HU ([-1.1, -1] %) |

Figure 3A-B demonstrate comparison of CT numbers acquired with the body vs. head phantom following the consensus configuration on the upright and recumbent CT, respectively, to assess the potential impact of beam hardening due to phantom size. On the upright CT, CT numbers were substantially higher using the head phantom (ΔHU≥27) except for two lung inserts (ΔHU≤8 for ρ=0.29-0.45 g/cm$^3$) with the maximum difference at the densest insert (ΔHU=136 at ρ=1.93 g/cm$^3$). Wilcoxon signed-rank test suggested statistically significant differences ($p<0.05$) between upright CT number quantifications using the body vs. head phantom for all inserts. On the recumbent CT, CT numbers agreed well between the two phantom sizes for soft tissue/adipose inserts (|ΔHU|≤8 for ρ=0.96-1.08 g/cm$^3$) while more discrepancies were observed for lung (max. ΔHU=-28 at ρ=0.29 g/cm$^3$) and bone (max. ΔHU=148 at ρ=1.93 g/cm$^3$). Supplement Table S3 summarizes the detailed CT numbers of above comparison. Figure 3C-D compare SPR HLUTs derived from a single session with date corresponding to the anthropomorphic phantom scan in thorax region using the body vs. head phantom following the consensus configuration on the upright and recumbent CT, respectively. For the upright CT, HLUTs showed substantial deviations around adipose/soft tissue (local maximum ΔSPR=0.049 a.u. (5.3%) at -136 HU) and cortical bone (maximum ΔSPR=0.081 a.u. (4.4%) at 1500 HU). For the recumbent CT, HLUTs also showed substantial deviations around cortical bone (maximum ΔSPR=0.084 a.u. (4.6%) at 1500 HU) and some deviations around lung (local maximum ΔSPR=0.020 a.u. (86.7%) at -950 HU).

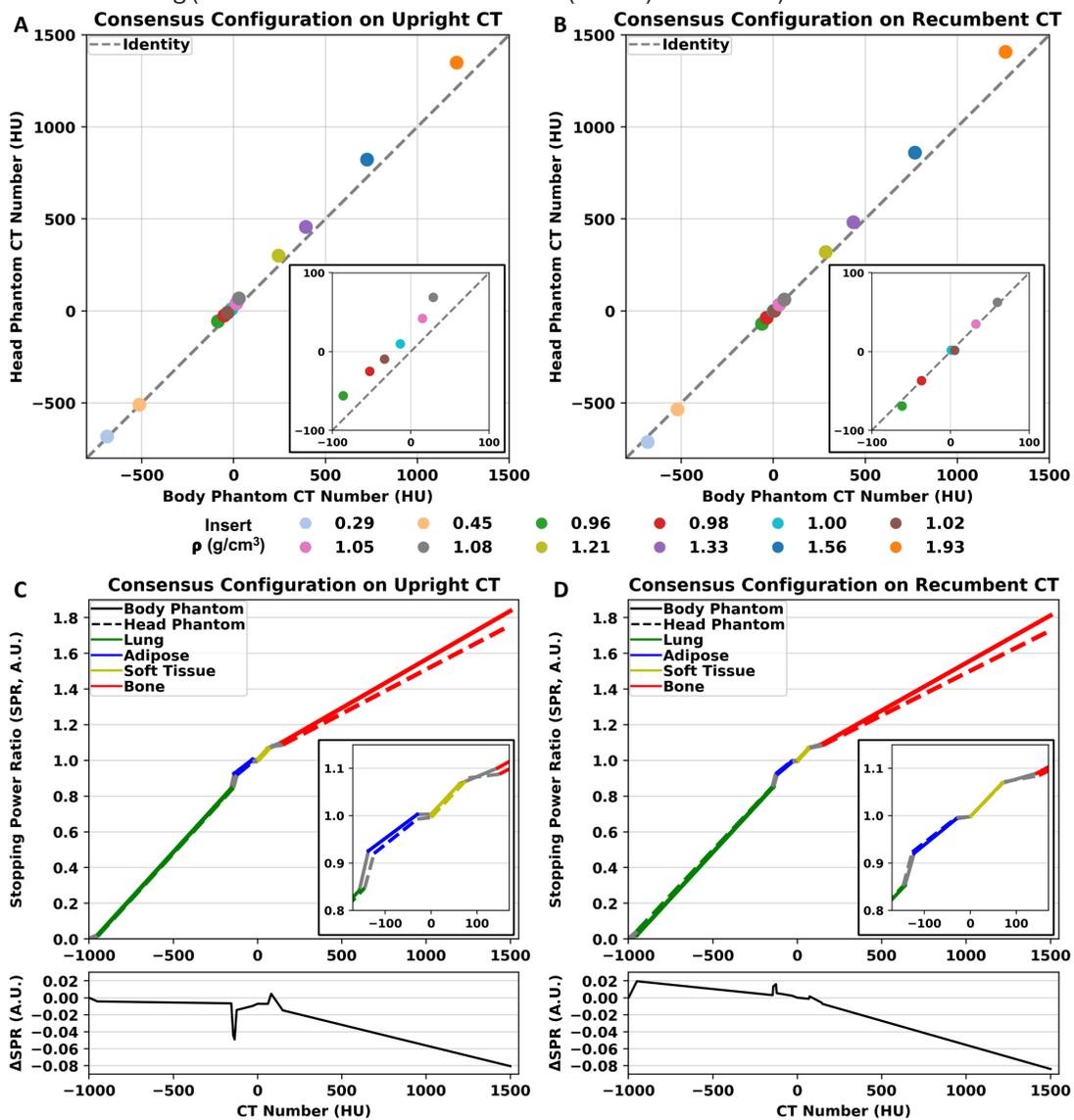

**Figure 3.** Comparison of acquisitions using the body versus head phantom following the consensus configuration to assess the potential impact of phantom size, including (A) CT numbers on the upright CT and (B) the recumbent CT, and (C) resultant Hounsfield look-up tables and their differences on the upright CT and (D) the recumbent CT. A zoomed-in view highlighting adipose and soft tissue is shown in the bottom right for each subfigure.

**Table S3.** Comparison of CT numbers acquired with the body versus head phantom following the consensus configuration on the upright and recumbent CT, along with Wilcoxon signed-rank test results detecting significant differences between CT number quantifications using the body versus head phantom on the upright CT.

| Configuration | Consensus configuration | | | | | |
|---|---|---|---|---|---|---|
| Machine | Upright CT (N=14) | | | Recumbent CT (N=1) | | |
| Metric | Body phantom Mean±SD (HU) | Head phantom Mean±SD (HU) | Difference in Mean (HU (%)) | Body phantom (HU) | Head phantom (HU) | Difference (HU (%)) |
| Lung LN300 | -689±1.1 | -681±1.8 | 8 (-1%) | -683 | -711 | -28 (-4%) |
| Lung LN450 | -514±1.1 | -509±0.9 | 6 (-1%) | -521 | -535 | -14 (-3%) |
| HE General Adipose | -86±2.0 | -56±1.9 | 30 (-35%) | -62 | -69 | -8 (-12%) |
| HE Breast 50:50 | -52±1.9 | -25±2.0 | 27 (-52%) | -37 | -37 | 0 (0%) |
| Liquid Water | -13±2.1 | 10±1.7 | 23 (-174%) | 1 | 2 | 1 (92%) |
| HE CT Solid Water | -33±2.3 | -10±2.0 | 24 (-71%) | 5 | 2 | -4 (-67%) |
| HE Brain | 15±2.1 | 42±1.8 | 27 (180%) | 32 | 35 | 3 (9%) |
| HE Liver | 29±2.3 | 69±1.9 | 40 (141%) | 59 | 63 | 3 (6%) |
| HE Inner Bone | 244±2.8 | 300±2.6 | 56 (23%) | 285 | 320 | 35 (12%) |
| CaCO3 30% | 392±3.0 | 458±3.0 | 66 (17%) | 434 | 483 | 49 (11%) |
| CaCO3 50% | 726±3.5 | 824±4.1 | 98 (14%) | 770 | 861 | 91 (12%) |
| HE Cortical Bone | 1213±4.9 | 1350±6.1 | 136 (11%) | 1261 | 1409 | 148 (12%) |
| Wilcoxon signed-rank test suggested statistically significant differences between quantifications of upright CT numbers using the body versus head phantom for *all* inserts (p<0.05) | | | | | | |

Supplement Table S4 summarizes CT number variations of bone inserts when placed off-center to further assess the impact of beam hardening. On both machines, CT numbers were generally lower when placed in the periphery of the phantom compared with in the center for all four bone inserts, while the differences were slightly higher on the upright CT (4-6%) compared with the recumbent CT (2-4%).

**Table S4.** Evaluation of off-center CT number variations of bone inserts using the body size phantom following the consensus phantom configuration on the upright and recumbent CT.

| Configuration | Consensus configuration | | | | | |
|---|---|---|---|---|---|---|
| Phantom size | Body | | | | | |
| Machine | Upright (N=14) | | | Recumbent (N=1) | | |
| Metric | Centered Mean±SD (HU) | Off-center Mean±SD (HU) | Difference in mean (HU (%)) | Centered (HU) | Off-center (HU) | Difference (HU (%)) |
| HE Inner Bone | 244±2.8 | 235±2.6 | -9 (-4%) | 285 | 272 | -12 (-4%) |
| CaCO3 30% | 392±3.0 | 375±3.0 | -17 (-4%) | 434 | 418 | -16 (-4%) |
| CaCO3 50% | 726±3.5 | 683±3.6 | -43 (-6%) | 770 | 748 | -21 (-3%) |
| HE Cortical Bone | 1213±4.9 | 1142±5.1 | -71 (-6%) | 1261 | 1241 | -20 (-2%) |

Figure 4A-B demonstrate comparison of CT numbers acquired following the consensus vs. simplified configuration using the body phantom on the upright and recumbent CT, respectively, to assess the feasibility of simplified configuration for routine measurements. On the upright CT, CT numbers were generally comparable between the two configurations with the largest absolute difference in the densest bone insert (Δ=-76 HU (-6%)). Wilcoxon signed-rank test suggested statistically significant differences (p<0.05) between upright CT number quantifications using the consensus vs. simplified configuration for all inserts. On the recumbent CT, the agreement is excellent and overall better compared to the upright CT, with largest absolute difference in the same densest bone insert (Δ=-23 HU (-2%)). Supplement Table S5 summarizes the detailed CT numbers of above comparison. Figure 4C-D compare SPR HLUTs derived from a single session with date corresponding to the anthropomorphic phantom scan in thorax region following the consensus vs. simplified configuration using the body phantom on the upright and recumbent CT, respectively. For the upright CT, HLUTs agreed well except for noticeable deviations around adipose/soft tissue (local maximum ΔSPR=0.014 a.u. (1.6%) at -136 HU) and cortical bone (maximum ΔSPR=0.047 a.u. (2.6%) at 1500 HU). For the recumbent CT, HLUTs showed excellent agreement globally with maximum ΔSPR=0.013 a.u. (0.7%) that occurred at 1500 HU.

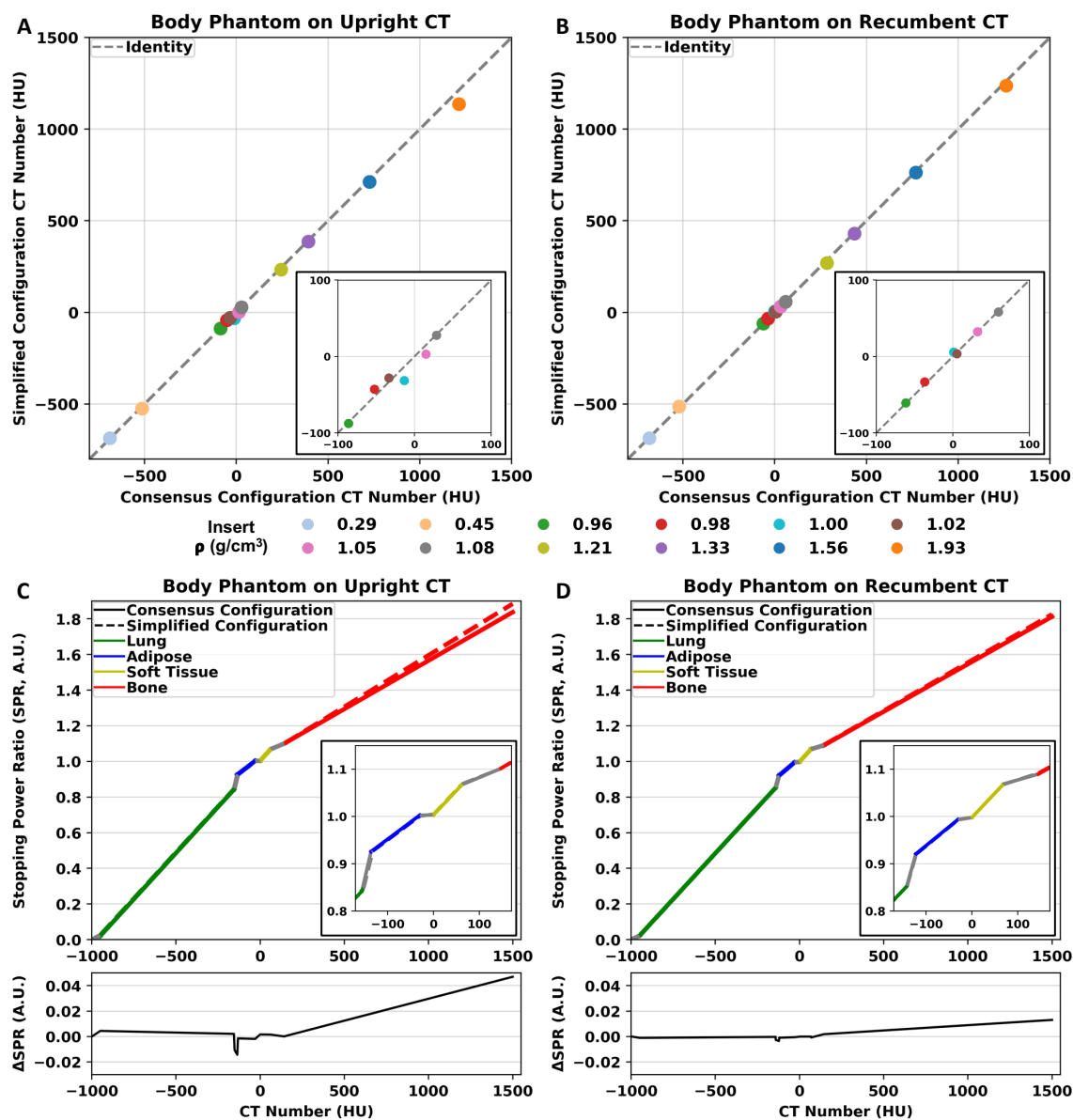

**Figure 4.** Comparison of acquisitions using the consensus versus simplified configuration with the body phantom to demonstrate the potential of routine simplified configuration measurements, including (A) CT numbers on the upright CT and (B) the recumbent CT, and (C) resultant Hounsfield look-up tables and their differences on the upright CT and (D) the recumbent CT. A zoomed-in view highlighting adipose and soft tissue is shown in the bottom right for each subfigure.

**Table S5.** Comparison of CT numbers acquired using the consensus versus simplified phantom configuration in the body phantom on the upright and recumbent CT, along with Wilcoxon signed-rank test results detecting significant differences between CT number quantifications using the consensus versus simplified configuration on the upright CT.

| Phantom Size | Body | | | | | |
|---|---|---|---|---|---|---|
| Machine | Upright CT | | | Recumbent CT | | |
| Metric | Consensus (N=14) Mean±SD (HU) | Simplified (N=15) Mean±SD (HU) | Difference in Mean (HU (%)) | Consensus (N=1) (HU) | Simplified (N=1) (HU) | Difference (HU (%)) |
| Lung LN300 | -689±1.1 | -686±0.8 | 3 (0%) | -683 | -687 | -4 (-1%) |
| Lung LN450 | -514±1.1 | -525±1.2 | -10 (-2%) | -521 | -513 | 8 (1%) |
| HE General Adipose | -86±2.0 | -88±2.1 | -2 (-2%) | -62 | -61 | 1 (1%) |
| HE Breast 50:50 | -52±1.9 | -43±1.8 | 9 (18%) | -37 | -33 | 4 (10%) |
| Liquid Water | -13±2.1 | -31±2.4 | -18 (-136%) | 1 | 6 | 5 (533%) |
| HE CT Solid Water | -33±2.3 | -28±2.0 | 5 (15%) | 5 | 4 | -2 (-33%) |
| HE Brain | 15±2.1 | 3±1.8 | -12 (-79%) | 32 | 32 | 0 (1%) |
| HE Liver | 29±2.3 | 28±1.9 | -1 (-3%) | 59 | 58 | -1 (-2%) |
| HE Inner Bone | 244±2.8 | 233±2.6 | -11 (-5%) | 285 | 269 | -16 (-5%) |
| CaCO3 30% | 392±3.0 | 387±2.4 | -5 (-1%) | 434 | 430 | -4 (-1%) |
| CaCO3 50% | 726±3.5 | 712±3.2 | -13 (-2%) | 770 | 764 | -6 (-1%) |
| HE Cortical Bone | 1213±4.9 | 1137±3.2 | -76 (-6%) | 1261 | 1237 | -23 (-2%) |
| Wilcoxon signed-rank test suggested statistically significant differences between quantifications of upright CT numbers using the consensus versus simplified configuration for *all* inserts (p<0.05) | | | | | | |

Figure 5 demonstrates dosimetric results for an anthropomorphic phantom comparing the spine plan dose calculated on different CT datasets and corresponding and HLUTs. Dose difference maps and line dose difference profiles demonstrated local differences up to 1.9 Gy (6.9% of prescription dose) comparing upright CT dose calculated using different HLUTs (1st row), while the differences were minimal comparing recumbent CT dose calculated using different HLUTs (2nd row). Local differences up to 1.8 Gy (6.6% of prescription dose) were also found comparing upright and recumbent CT dose both calculated using consensus-derived HLUTs (3rd row). Nevertheless, clinical DVH curves suggested excellent agreement with minimal target coverage differences with $\Delta D_{95}$<0.2% and gamma pass rates were 100.0% for all comparisons.

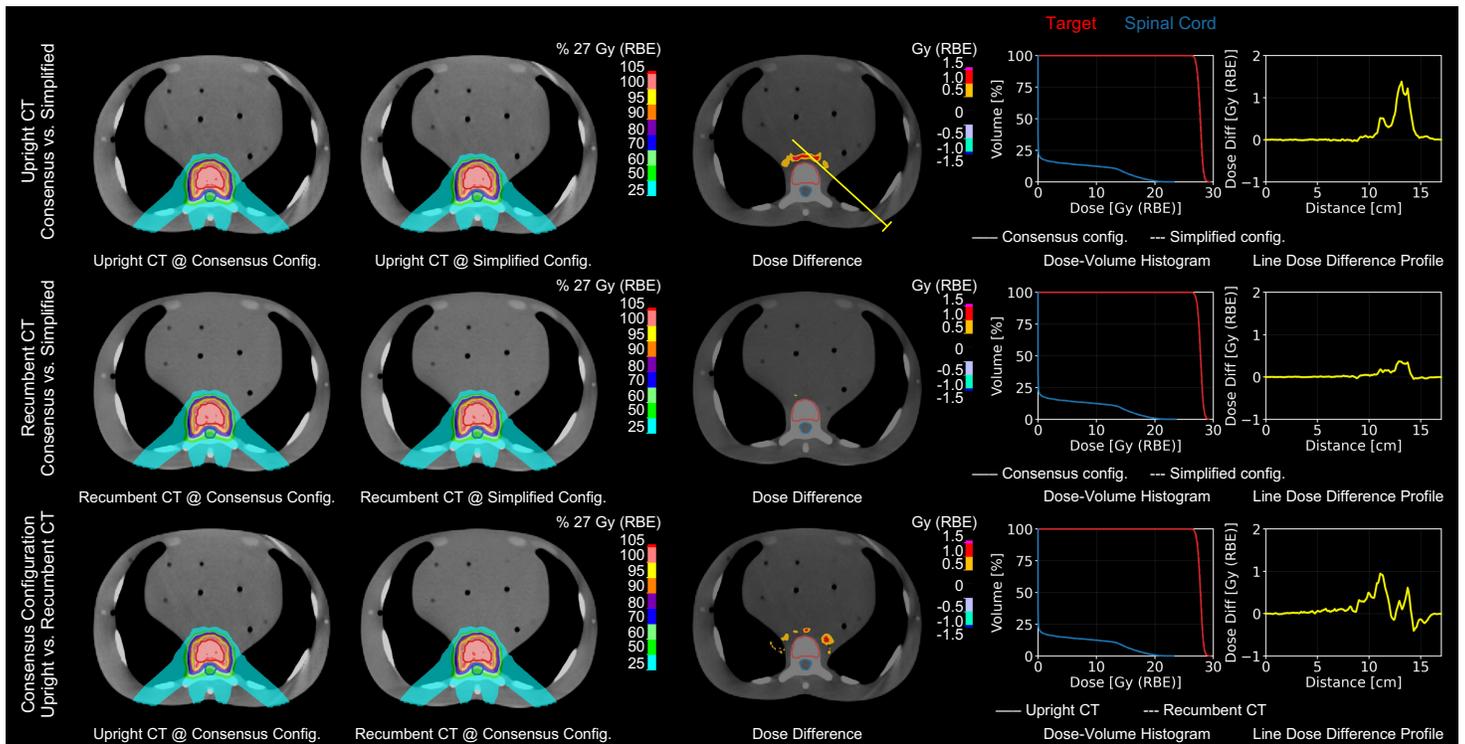

**Figure 5.** Dosimetric comparison of a spine plan calculated on different CT dataset and/or Hounsfield look up tables demonstrating dose distribution and comparison of dose, dose-volume histograms and line dose profiles.

Figure 6 demonstrates the dosimetric impact of CT and HLUT on a prostate cancer plan. Similar to the spine plan, as suggested by dose difference maps and line dose difference profiles, local dose differences of up to 2.0 Gy (5.0% of prescription dose) were present when comparing upright CT dose calculated using different HLUTs (1st row). The dose differences in phantom configuration were less pronounced when comparing recumbent CT dose calculated using different HLUTs (2nd row). Local differences up to 3.0 Gy (7.5% of prescription dose) on the proximal edge of the pelvic bones were observed between the upright and recumbent CT dose calculated using their respective consensus-derived HLUTs (3rd row). Yet, despite these local differences, the corresponding DVH curves suggest excellent agreement with minimal target coverage differences ($\Delta D_{98}$<0.1% for all comparisons), and gamma pass rates were 100.0% for all comparisons.

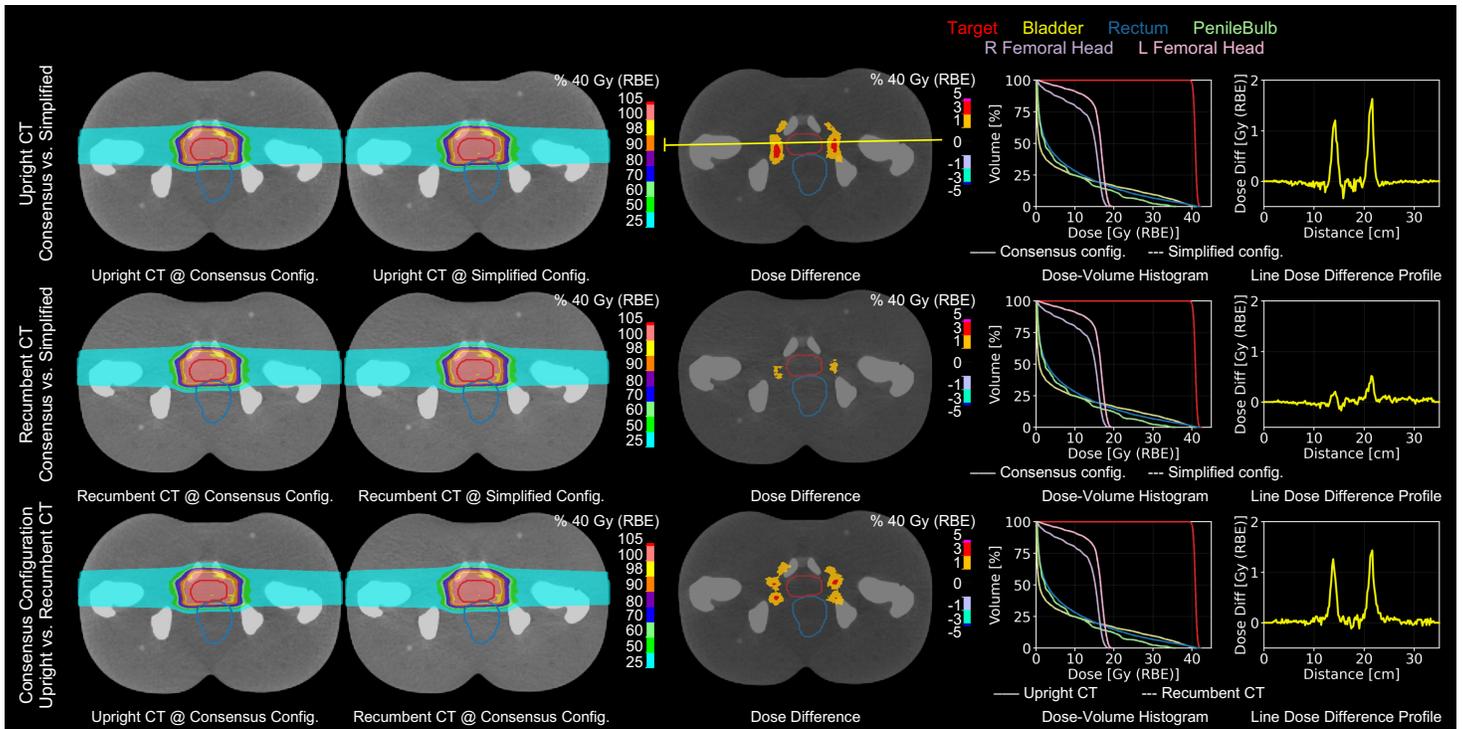

**Figure 6.** Dosimetric comparison of a prostate plan calculated on different CT dataset and/or Hounsfield look up tables demonstrating dose distribution and comparison of dose, dose-volume histograms and line dose profiles.

## Discussion

Our work presents the first longitudinal assessment of upright CT stability over an 8-month period. Data were obtained following consensus guidelines[13] and using a simplified phantom configuration to evaluate feasibility for establishing a routine QA program. For both consensus and simplified configurations, upright CT number showed excellent repeatability with intra- and inter-session SD <5 HU. These results were comparable to reported longitudinal performance of a recumbent CT where CT number SD=0.9-6.4 HU for 100 repeat measurements of materials of a Catphan phantom (Phantom Laboratory, Salem, NY) over 6 months,[19] suggesting equivalent longitudinal stability of upright CT to standard of care recumbent CT. These results are meaningful in the context of offline and online adaptive proton therapy, where daily IGRT images may be considered for QACT evaluation. Given that ~30-60% head and neck cancer patients and ~60% thoracic cancer patients are reported to require adaptation in proton therapy[20–23], having the ability to use isocentric, high quality upright CT data for offline replanning in lieu of QACTs is advantageous. Implementing upright CT for routine online or offline ART would enable efficient and reliable dose verification with accurate patient anatomy in the treatment position, thereby eliminating uncertainties from deformable registration which would reduce clinical resources and accelerate the clinical ART workflow[20,21].

During the 8-month study, the upright CT underwent software upgrade twice and recalibration once during one of the software upgrades. Systematic offsets were found after the first software upgrade with simultaneous recalibration, which remained consistent after the second upgrade. Following standards for IGRT and CT Simulator QA[11,24], post-upgrade verifications and potential re-baselining the SPR HLUT is expected after each major system upgrade.

Comparing acquisitions using the body and head phantoms, large discrepancies were seen especially around cortical bone for CT number measurements (up to 148 HU (12%)) and SPR derivations (up to 0.084 a.u. (5%)) on both CT machines. Literature also reported discrepancies between SPRs derived from the body and head phantom acquisitions for the same vendor as our recumbent CT (0.019 a.u. at 2000 HU) which was higher compared with other vendors (<0.006 a.u. at 2000 HU).[13] Our experiments suggest that site-specific HLUTs are needed to accommodate the size-dependent discrepancies in SPR calibration. Alternatively, efforts could be made towards reducing such discrepancies at the imaging stage, for example, employing additional beam hardening correction during reconstruction or postprocessing.[25,26]

Comparing acquisitions following the consensus and simplified configurations, CT numbers generally agree except for higher discrepancies in bone (up to 76 HU) especially on the upright CT. Resultant SPR demonstrated discrepancies <2.6% on the upright CT and <0.7% on the recumbent CT. Ainsley et al.[18] performed a study in a manner similar to ours by comparing SPR acquired by scanning all inserts at once vs. one by one (similar to 'simplified' vs. 'consensus' in our experiments) and obtained ΔSPR<1% between the two methodologies, which are on similar magnitude to our findings. They also observed correlation of increasing SPR difference with higher HU in bone region (~HU>200) that also agreed with our findings. Furthermore, on both upright and recumbent CT, dose calculation in complex phantom geometries including heterogeneous tissue (e.g., bone, lung, tissue interfaces) showed global agreement with local differences (up to 8% of prescription dose) when using HLUTs derived from different phantom configurations, suggesting that proton dose calculation is modestly sensitive to the potential uncertainties in calibration contributed by the simplified phantom configuration, including scatter, beam hardening, etc.[27]

Despite clinically acceptable dosimetric agreement in our experiments (100% gamma pass rate and <0.2% difference in target coverage), the simplified configuration is not intended for derivation of clinical HLUT for dose calculation purposes and instead is presented as a means to consider for routine QA. Due to the configuration, more beam hardening uncertainty is introduced which could potentially lead to different HU numbers,[18] for example ΔHU=-76 HU for one dense insert in our experiment. However, the faster acquisition using a single scan renders this approach suitable for more frequent, routine QA to monitor the machine stability to support online proton ART, as its longitudinal variability showed minimal differences to scans following the consensus guidelines (simplified vs. consensus intersession SD <3 HU vs. < 5 HU).

One limitation of this work is that the experiments were only performed in phantom whereas further in-vivo/ex-vivo validations are needed. Besides validation in real patient cohort, one of the common approaches is to experimentally measure the water equivalent length of proton using representative animal organs (soft tissue, adipose, bone, lungs) and compare against CT derivations[28,29], which are beyond the scope of the current reliability study and phantom work.

## Conclusion
The upright CT demonstrated excellent longitudinal stability and reliability to support proton ART. The simplified configuration shows feasibility to serve as rapid QA to monitor machine stability. With further validation across multiple institutions, online upright proton ART can be further established.